# Percolative Effects on Noise in Pentacene Transistors

B. R. Conrad, W. G. Cullen, W. Yan, E. D. Williams

Department of Physics, University of Maryland, College Park, College Park MD 20742

Noise in pentacene thin film transistors has been measured as a function of device thickness from well above the effective conduction channel thickness to only two conducting layers. Over the entire thickness range, the spectral noise form is 1/f, and the noise parameter varies inversely with gate voltage, confirming that the noise is due to mobility fluctuations, even in the thinnest films. Hooge's parameter varies as an inverse power-law with conductivity for all film thicknesses. The magnitude and transport characteristics of the spectral noise are well explained in terms of percolative effects arising from the grain boundary structure.



Despite significant interest in the performance of organic thin film transistors, transport mechanismsremain poorly understood. Of particular importance is the issue of conduction channel quality. As organic devices decrease in size, their active conduction channels become thinner and the signal to noise ratio is expected to increase[1]. However, conductance noise is generally seen to be more sensitive than the absolute conductance to conduction channel defects [2]. Therefore the importance of understanding noise mechanisms is an increasing concern[3 4 5, 6]. Previous reports show that mobility fluctuations generate noise in organic devices, but reports of the noise magnitude and dependence on device parameters such as gate voltage $V_g$ and source-drain voltage $V_{sd}$ vary significantly[7-10].

We have measured the spectral drain current noise of Pn-TFTs as a function of the device organic conduction layer thickness to clarify the noise generation mechanism. Device thickness is studied from a thick film regime[11-14] of an electrostatically limited conduction channel to the thin-film regime where transport is physically limited to two continuous layers of pentacene as shown in Figure 1. It is found that slow-grown Pn-TFTs display 1/f noise that is independent of conduction channel thickness and that the main mechanism of noise generation is consistent with mobility fluctuations over the entire thickness range. The noise magnitude increases with decreasing mobility. It is shown that these dependencies can be quantified in terms of changes in the channel conductance similar to those of percolative systems.

Pentacene device I/V characteristics are highly sensitive to the substrate surface chemistry and topography, since these properties greatly influence the first few layers of the thin film and therefore the majority of the conduction channel[11, 13, 15]. Here, pentacene was thermally deposited on heavily doped silicon (100) wafers with 300 nm thermally grown oxide, with growth temperature 330K and a deposition rate 0.03Å/s. The top-contact devices had 50nm thick



Au source-drain electrodes deposited *in-situ*. Transport measurements were conducted in a nitrogen atmosphere at room temperature after at least a 12 hour degassing period to minimize atmospheric doping. All reported results are averaged over measurements of at least three separate devices. The averaged transfer curves for films ranging in thickness from 40nm to 7nm and a set of 10nm devices with a deposition rate of 3.6 Å/s are shown in Figure 2a. As has been reported previously [16], an optimum mobility is obtained for intermediate thicknesses of 15-25 nm.

Organic thin film transistors (TFT's) typically display flicker (1/f) noise[7, 9] in the linear IV regime for low frequencies of roughly 1Hz-10kHz[2, 17]. Empirically, we expect the current noise to have the form

$$S = A \cdot I_{SD}^2 \big/ f^{\alpha} \qquad (1)$$

where S is the spectral noise density, f is the frequency, the exponent $\alpha$ is a constant, A is the noise magnitude coefficient, and $I_{SD}$ is the device source drain current. For all the devices prepared as described above, the spectral noise density was observed to obey

$$S \propto I^{2.0 \pm 0.1} \big/ f^{1.0 \pm 0.1} \qquad (2)$$

for the range 3Hz – 10kHz, as shown with the lower curve in Fig. 2b. Deviations from the typical $\alpha = 1$ behavior for organics usually result from a far from equilibrium growth rate[18, 19] as can be seen in the upper curve in Figure 2b.

When the noise is primarily due to mobility fluctuations, the noise coefficient can be written as $A = \alpha_H / N$ for homogeneous conduction channels under homogeneous electric fields, where $\alpha_H$ is Hooge's parameter and N is the total number of carriers in the system[1, 2, 20]. The number of charge carriers is well estimated by treating the device as a parallel plate capacitor:



$$N = \frac{c_g}{e} \cdot (V_g - V_{th} - V_{sd}/2) \ , \hspace{2cm} (3)$$

where $c_g = L \cdot W \varepsilon_{SiO2} \cdot \varepsilon_{Pn} / (d_{SiO2} \cdot \varepsilon_{Pn} + d_{Pn} \cdot \varepsilon_{SiO2})$, and $V_{SD}$, $V_G$ and $V_{th}$ are the source-drain, gate, and threshold voltages, $c_g$ is the capacitance, and $\varepsilon_{SiO2}$ ($d_{SiO2}$) and $\varepsilon_{Pn}$ ($d_{Pn}$) are the dielectric constants (thicknesses) of the insulating layer and pentacene. Within the linear I/V regime, the noise coefficient can then be written as

$$A = \frac{\alpha_H \cdot e}{c_g} \cdot \frac{1}{(V_g - V_{th} - V_{sd}/2)} \hspace{1cm} (4)$$

It can be seen in Figure 3 that the measured noise coefficient A is inversely proportional to the effective conduction channel voltage $(V_g - V_{th} - V_{sd}/2)$ , consistent with expectations for a homogenous semiconductor governed by mobility fluctuations[1, 2, 21]. The value of Hooge's parameter can be determined from the noise coefficient A, using Eq. 4 as a function of the conduction channel thickness, as in Figure 4a. For thicknesses below 10nm, physical limitations on the conduction channel, and therefore mobility, are expected[11]. The thickness-dependent mobility is shown in Fig. 4b. Figs. 4a and 4b together show that Hooge's parameter has a strong dependence on mobility, specifically $a_H \propto 1/\mu^w$ where $w = 2.9 \pm 0.4$ . The large values of Hooge's parameter at small film thicknesses are consistent with previous observations for inhomogeneous samples[20].

An increase in noise levels in inverse proportion to mobility is expected in the case of semiconductor transport limited by impurity scattering[2], and strong correlations of noise with mobility are often observed in devices withlimited mobility[3, 17, 22]. Increased noise in thin films has also been attributed to current crowding in inhomogeneous films [20, 23]. It is generally accepted that organic semiconductor transport is dominated by hopping and that grain boundaries



disrupt transport, despite debate as to the detailed mechanism[24]. Therefore, it is reasonable to

expect analogies with the behavior of noise in disordered materials, where a percolative model

can be applied. Consider that resistivity ρ is inversely proportional to the number of carriers N,

and the mobility, $\rho \propto 1/N \cdot \mu$. If we fix the number of charge carriers N with a gate voltage, the

mobility is the determinant of resistivity. In a percolation model the sample is treated as a

mixture of conducting and insulating components. As the fraction of the conducting component

$p$ increases beyond some critical fraction $p_c$, the sample resistivity exhibits a power law decay of

the form $\rho \propto (p - p_c)^{-t}$. The corresponding scaled current noise S/I$^2$ is observed to decay as

$S/I^2 \propto (p - p_c)^{-\kappa}$, where κ is a model dependent-parameter[25]. From the expressions for the scaled

current noise density $S/I^2$ and the resistivity $\rho$, the scaled current noise depends on channel

resistivity as $S/I^2 \propto \rho^w$, and thus at fixed number density as $S/I^2 \propto \mu^{-w}$, with the exponent $w=k/t$.

For the devices described here, $w = 2.9 \pm 0.4$ is in agreement with the value of $w$ previously

calculated and different percolation models[4]. It is important to note that these power law

behaviors are only universal near percolative threshold $p_c$, even though effective power law

behavior is often observed over a wide range. The essential insight to be gained from the

connection to percolation models is that the Pn-TFT's noise is correlated with the distribution of

the conducting elements which could also cause current crowding and concomitant increases in

noise[20, 26]. As seen in Figure 1, the most prominent defect in a pentacene thin film is a grain

boundary and therefore the most likely mechanism for the mobility dependence of Hooge's

parameter is transport across these boundaries.

Even though the I-V characteristics of Pn-TFT's are highly sensitive to the dielectric

interactions in the bottom-most layers, the limitation of conduction to only a few layers of

pentacene does not change the functional form of the noise behavior. Therefore, the primary 1/f



noise mechanisms must be similar over the entire conduction channel thickness. The results are consistent with conductivity fluctuations due to charge hopping through the resistive barriers[27] between grains. This mechanism is likely to explain similar noise signatures seen in organic polymers[28] as well. Interlayer transport effects might contribute to the noise but the anisotropic mobility for these types of organic molecules[29, 30] and the layer-dependent measurements reported here suggest that interplanar effects are small in comparison.

In conclusion, mobility fluctuations dominate the 1/f noise over the full range of pentacene film thickness from two continuous layers to an electrostatically limited conduction channel. Large values of Hooge's parameter are explained by quantifying their variation as a power-law with the conductivity, similar to the behavior of percolative systems and suggesting the importance of the random spatial distribution of grain boundaries to the noise generation mechanism. Surprisingly, the functional form of the resistive grain boundary-generated noise is independent of proximity to the gate oxide.

**Acknowledgments:** This work was supported by NIST under contract #70NANB6H6138, by the LPS, and by the UMD NSF-MRSEC under grant DMR 05-20471. We are grateful to M. Ishigami and D. Dougherty for setting up the noise measurements and for extremely insightful discussions.

Figures

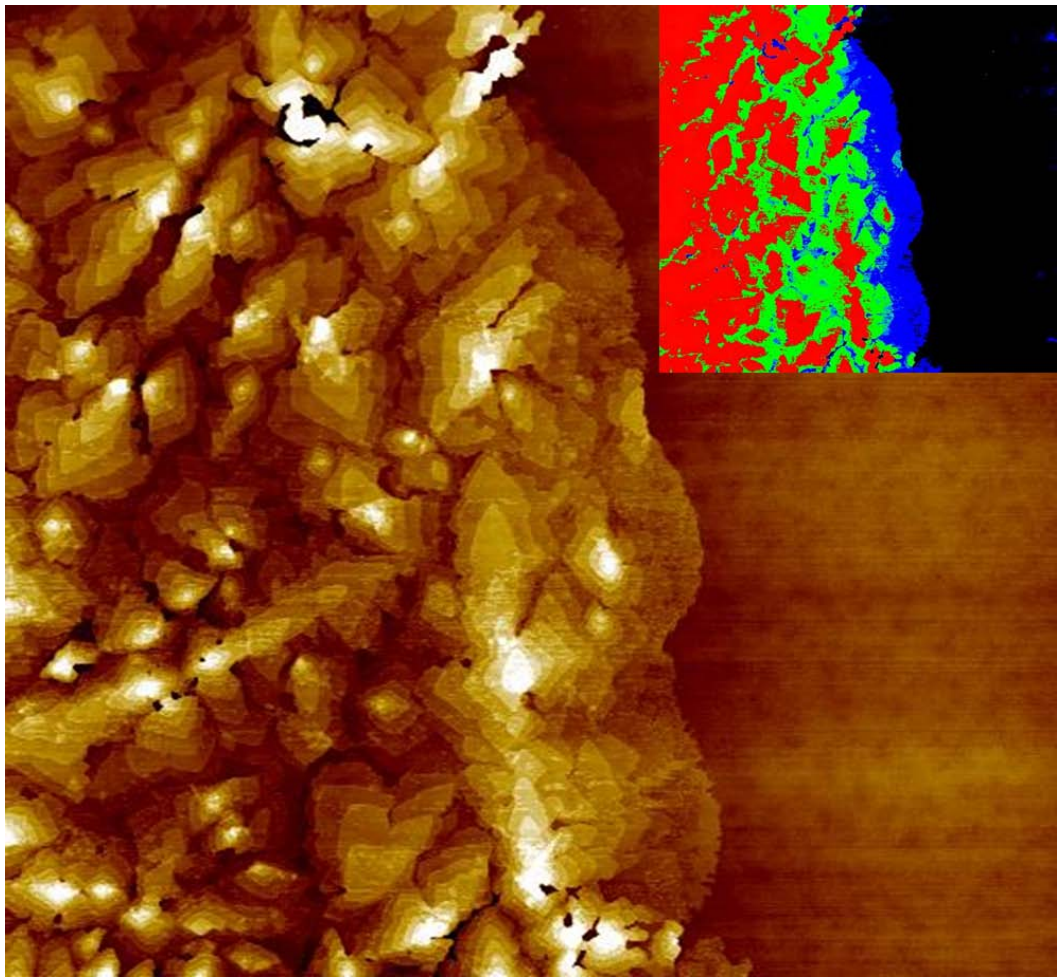

**Figure 1:** (color on line)  A $(20um)^2$ Atomic Force Microscopy image of the conduction channel edge of a 7nm pentacene device. Three complete pentacene layers are mostly covered by increasingly less complete layers. The inset depicts the same image with shading (colors) representing mostly complete pentacene layers where dark gray (blue) is the first pentacene layer, white (green) is the second and third pentacene layers, and medium gray (red) is larger than three layers.



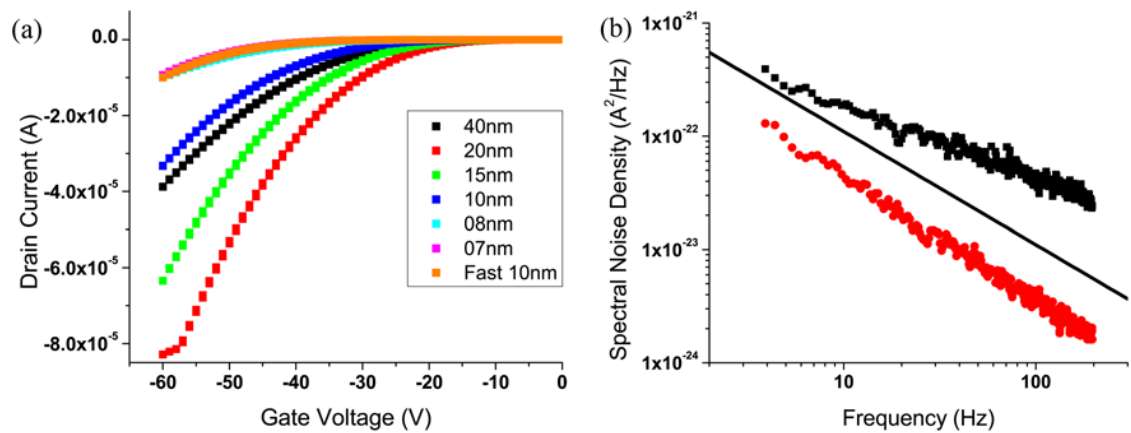

**Figure 2:** (color on line) **(a)** Averaged transfer curves for film thickness 40nm to 7nm and a set of 10nm devices fabricated with a deposition rate 120 times **(b)** Spectral noise density S for 10 nm thick devices: open red circles - devices fabricated with slow pentacene growth, $S \propto f^{-1.05}$; black squares - devices fabricated with fast pentacene growth, $S \propto f^{-0.64}$. The black line represents the theoretical $S \propto 1/f$.





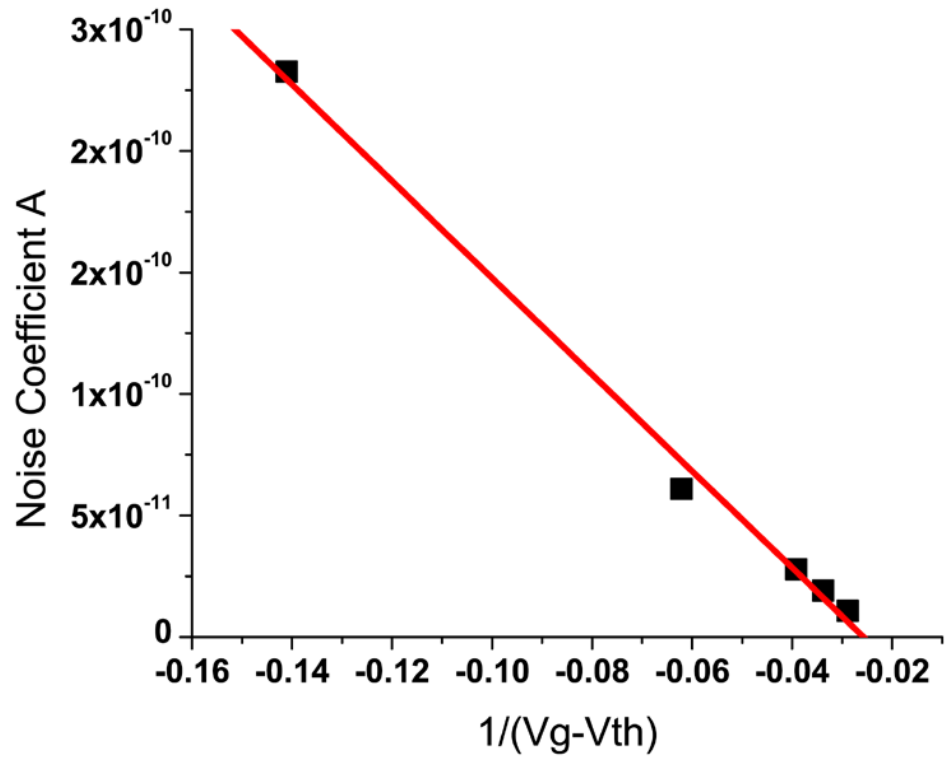

**Figure 3:** (color on line) Noise constant (Eq. 1) as a function of inverse difference between the gate voltage and the threshold voltage. The black squares indicate a typical device while the line indicates dependence $A \propto 1/(V_g - V_{th})$.



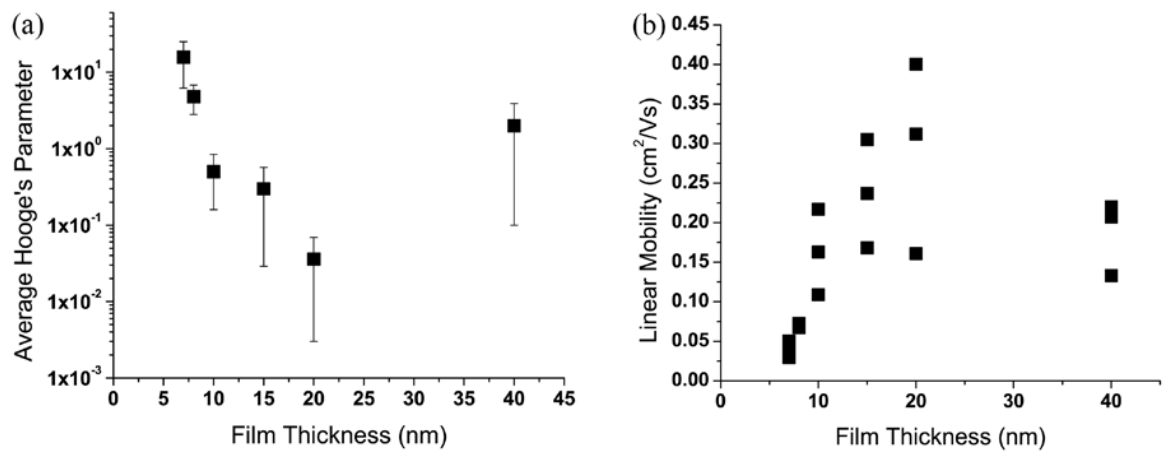

**Figure 4: (a)** Average Hooge's constants (Eq. 4) and **(b)** mobility as a function of the device pentacene film thickness.